\documentclass{PoS}

\title{NA61/SHINE at the CERN SPS}

\ShortTitle{NA61/SHINE at the CERN SPS}

\author{\speaker{Andras Laszlo} for the NA61 Collaboration\\
        KFKI Research Institute for Particle and Nuclear Physics, Budapest, Hungary\\
        E-mail: \email{laszloa@rmki.kfki.hu}}

\author{The NA61 Collaboration:\\
N.~Abgrall${}^{23}$,
A.~Aduszkiewicz${}^{24}$,
B.~Andrieu${}^{11}$,
T.~Anticic${}^{13}$,
N.~Antoniou${}^{18}$,
A.~G.~Asryan${}^{15}$,
B.~Baatar${}^{9}$,
A.~Blondel${}^{23}$,
J.~Blumer${}^{5}$,
L.~Boldizsar${}^{10}$
% ...
}

\author{\\
\vspace*{1cm}\\
${}^{ 1}$Cape Town University, Cape Town, South Africa \\
${}^{ 2}$ETH, Zurich, Switzerland \\
${}^{ 3}$Fachhochschule Frankfurt, Frankfurt, Germany \\
${}^{ 4}$Faculty of Physics, University of Sofia, Sofia, Bulgaria \\
${}^{ 5}$Forschungszentrum Karlsruhe, Karlsruhe, Germany \\
${}^{ 6}$Institute for Nuclear Research, Moscow, Russia \\
${}^{ 7}$Institute for Particle and Nuclear Studies, KEK, Tsukuba,  Japan \\
${}^{ 8}$Jagellionian University, Cracow, Poland  \\
${}^{ 9}$Joint Institute for Nuclear Research, Dubna, Russia \\
${}^{10}$KFKI Research Institute for Particle and Nuclear Physics, Budapest, Hungary \\
${}^{11}$LPNHE, University of Paris VI and VII, Paris, France \\
${}^{12}$Pusan National University, Pusan, Republic of Korea \\
${}^{13}$Rudjer Boskovic Institute, Zagreb, Croatia \\
${}^{14}$Soltan Institute for Nuclear Studies, Warsaw, Poland \\
${}^{15}$St. Petersburg State University, St. Petersburg, Russia \\
${}^{16}$State University of New York, Stony Brook, USA \\
${}^{17}$\'Swi{\,e}tokrzyska Academy, Kielce, Poland \\
${}^{18}$University of Athens, Athens, Greece \\
${}^{19}$University of Bari and INFN, Bari, Italy \\
${}^{20}$University of Bergen, Bergen, Norway \\
${}^{21}$University of Bern, Bern, Switzerland \\
${}^{22}$University of Frankfurt, Frankfurt, Germany \\
${}^{23}$University of Geneva, Geneva, Switzerland \\
${}^{24}$University of Warsaw, Warsaw, Poland \\
${}^{25}$Warsaw University of Technology, Warsaw, Poland  \\

${}^{ \#}$Now at University of Wisconsin, Madison, USA. \\}

\abstract{
Status of the new experimental program 
to study hadron production in hadron-nucleus and
nucleus-nucleus collisions at the CERN SPS will
be presented. In particular, a detailed physics motivation
and experimental strategy will be given for the part of the
program related to the physics of strongly interacting matter:
search for the critical point of strongly interacting matter,
study properties of the onset of deconfinement, and
high $p{{}_T}$ measurements in p+p and p+A interactions.
The planned measurements for the neutrino T2K and cosmic-ray
experiments will also be discussed.}

\FullConference{Critical Point and Onset of Deconfinement   -  4th International Workshop\\
		 July 9 - 13, 2007\\
		 Darmstadt, Germany}

\begin{document}

\section{Introduction}

The NA61/SHINE \cite{homepage} is a new fixed-target experiment 
at the CERN SPS accelerator, based on the upgraded setup of the 
NA49 apparatus. In particular, the most expensive components are inherited from the 
NA49. These are the two superconducting magnets, the four large volume TPCs and the 
two ToF walls.

The physics programme of NA61 is the systematic 
measurement of hadron production in proton-proton, 
proton-nucleus, hadron-nucleus, and nucleus-nucleus collisions. 
This comprehensive study mainly concerns the following objectives.
\begin{enumerate}
 \item Search for the critical point by an energy - system size 
       scan.
 \item Study the properties of the onset of deconfinement by the 
       energy - system size scan.
 \item To establish, together with the RHIC results, the energy dependence 
       of the nuclear modification factor.
 \item To record hadron-nucleus reference spectra for the T2K neutrino 
       experiment, and for the Pierre Auger Observatory and KASCADE 
       cosmic-ray experiments.
\end{enumerate}

The detector upgrades shall also be discussed in the light 
of the necessities, posed by the physics goals.

\section{Physics goals}

\subsection{Search for the critical point and onset of deconfinement}

As predicted by lattice QCD calculations (see e.g.\ \cite{katzfodor}), 
the phase diagram of the strongly interacting matter admits a phase 
border of 1-st order phase transition on the temperature - baryochemical 
potential, which has a critical endpoint. 
According to the calculations, this critical endpoint may be located 
in the energy range, accessible at the CERN SPS.

\begin{figure}[!ht]
\begin{center}
\includegraphics[width=7cm]{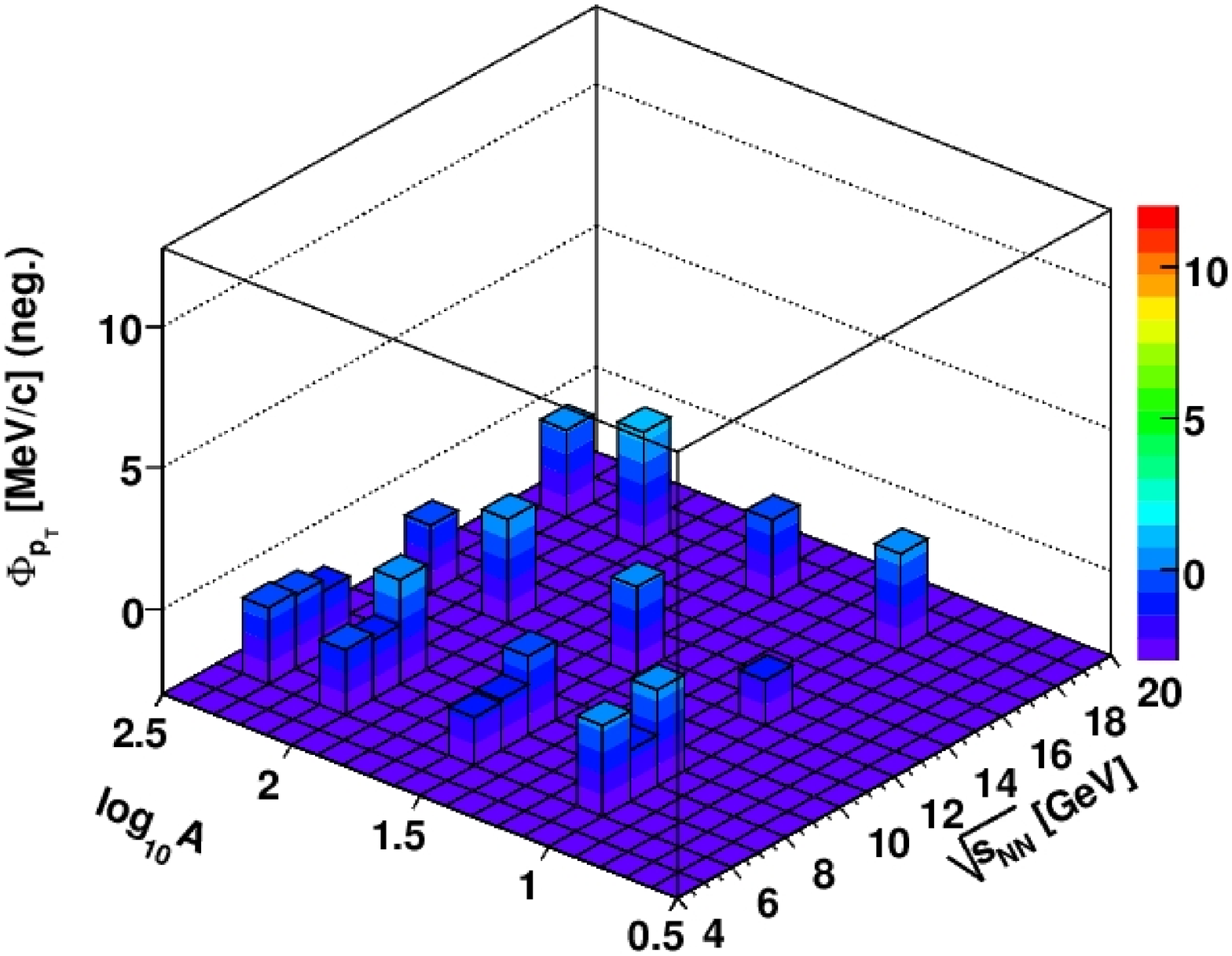}\includegraphics[width=7cm]{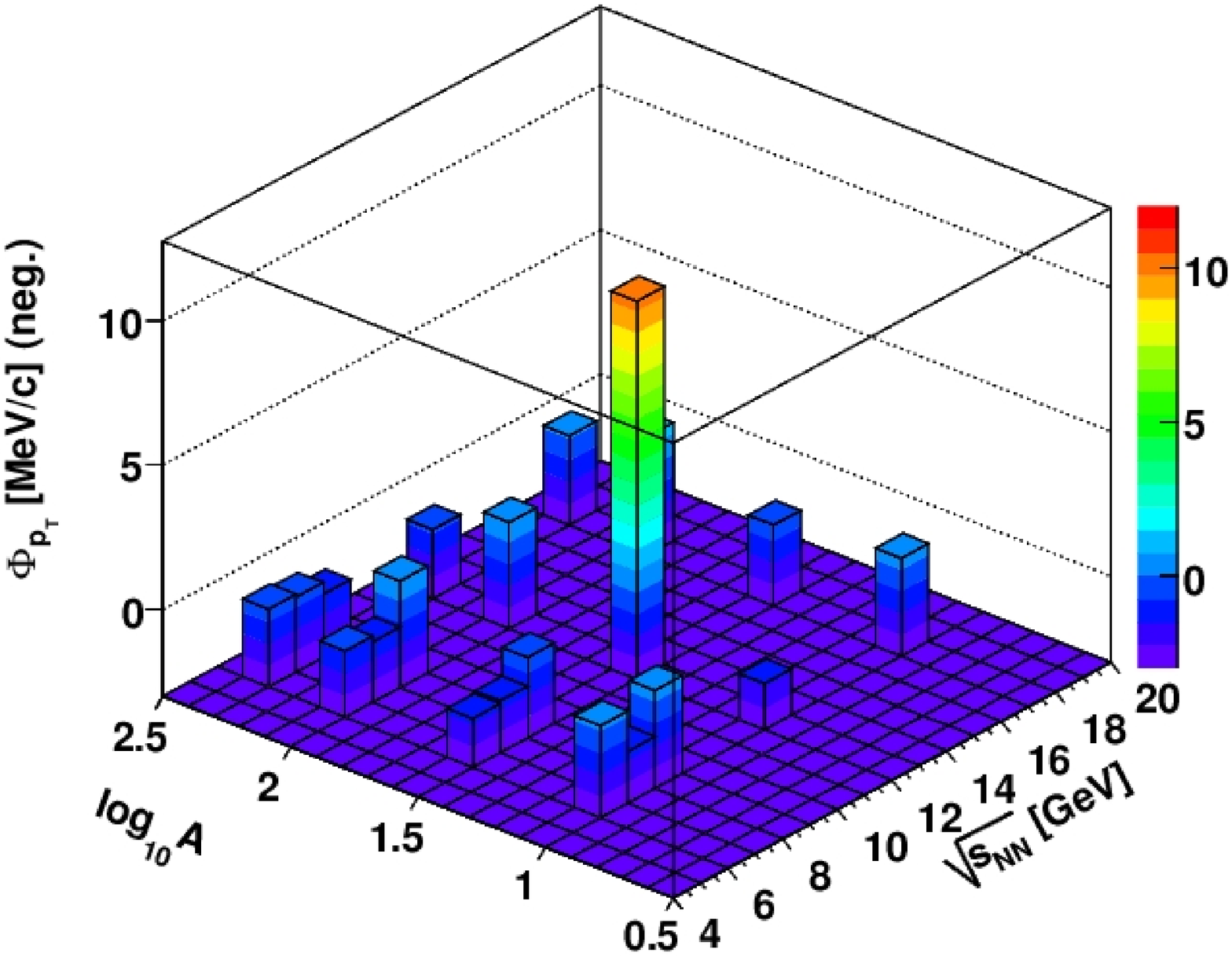}
\end{center}
\caption{Performance simulation for measuring transverse momentum fluctuations. 
UrQMD data (left) and artificial fluctuations added to this background (right).}
\label{fluct}
\end{figure}

Although the temperature ($T$) and baryochemical potential ($\mu_{B}$) 
are not directly measurable quantities, the $T - \mu_{B}$ coordinates 
of the freeze-out points of nuclear reactions can be brought into 
one-to-one correspondence with the energy ($E$) and system size ($A$) 
of the nuclear collisions (see e.g.\ \cite{becattini}). Therefore, 
the $T - \mu_{B}$ coordinates of the freeze-out points may be scanned 
via a systematic $E - A$ scan. When the phase evolution of a given 
collision passes near the critical endpoint, the increase of 
scaled variance ($\omega$) of multiplicity distribution and transverse 
momentum fluctuations measure ($\Phi_{p_{{}_T}}$) are expected 
(see e.g.\ \cite{rajagopal}). Thus, the critical endpoint may be 
discovered by looking at the energy and system size dependence of 
the multiplicity and transverse momentum fluctuations. A performance 
simulation of the detector for measuring transverse momentum 
fluctuations is shown in Figure~\ref{fluct}. 
The $\Phi_{p_{{}_T}}$ measure calculated within the UrQMD model in the 
NA61 acceptance is shown for the reactions planned to be taken 
by NA61 (left panel). In the right panel additional fluctuations 
($10\,\mathrm{MeV/c}$) as expected in the vicinity of the critical 
point are added to a point for S+S collisions at $80A\,\mathrm{GeV}$ \cite{KG}.

\begin{figure}[!ht]
\begin{center}
\includegraphics[width=10cm]{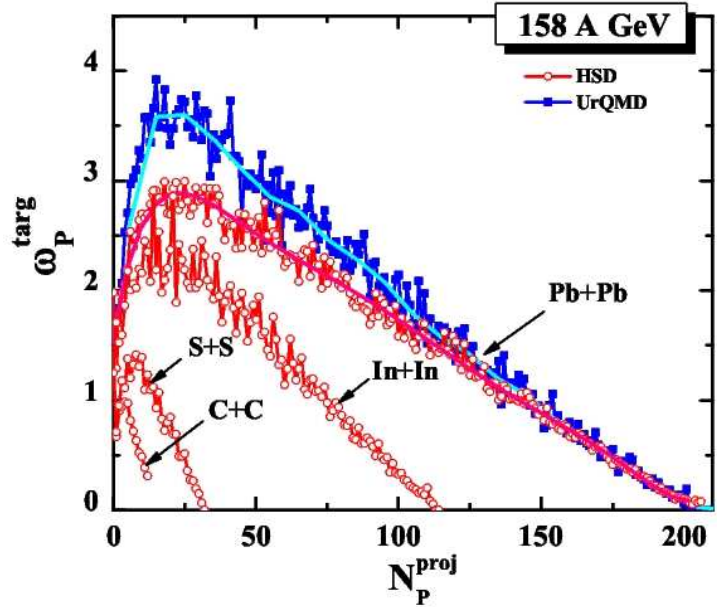}
\end{center}
\caption{Fluctuation of the number of target participants, as a function of projectile participants.}
\label{fluctbg}
\end{figure}

The detector upgrades and reactions to be studied 
are determined by the necessity of the reduction 
of the background fluctuations, in particular caused by the 
fluctuations of the number of participant nucleons. 
As shown in Figure~\ref{fluctbg} even for a fixed number of the projectile 
participants (by a calorimetric measurement of the number of 
projectile spectators) the number of target participants fluctuates \cite{konchakovski}. 
These fluctuations which constitute a significant background 
in a search for fluctuation signals of the critical point and the 
onset of deconfinement can be suppressed by selecting very central 
collisions of identical nuclei.
The present centrality trigger facility, 
the VCAL (a downstream calorimeter), is measuring the energy, 
carried by the projectile spectators: 
the collision is central if the spectator energy is small. However, 
the energy resolution of the VCAL is not good enough: a new projectile 
spectator energy measurement facility, the Projectile Spectator Detector 
(PSD), is being built, with much better resolution, to enable 
a precise selection of very central collisions.

An other source of background for multiplicity fluctuation measurements 
is the contamination by spiralling low-energy knock-on electrons 
($\delta$-electrons). To minimize this contribution, a Helium beam pipe 
will be introduced in the beam-line along the sensitive TPC volumes.

The planed systematic scan in energy and system size will 
allow to study the system size dependence of the anomalies 
in hadron production observed by NA49 \cite{Afanasiev:2002mx} in central Pb+Pb 
collisions at about 30A GeV. These anomalies were predicted 
for the onset of deconfinement \cite{Gazdzicki:1998vd} and their further understanding 
requires new NA61 data.

\subsection{Study high transverse momentum particle spectra}

A very interesting phenomenon, discovered by RHIC experiments at 
$\sqrt{s_{{}_{NN}}}=200\,\mathrm{GeV}$ collision energy is the reduction 
of high transverse momentum particle production in nuclear collisions 
relative to elementary collisions (see e.g.\ \cite{adler}), when assuming 
scaling of particle spectra by the number of binary collisions. This 
phenomenon is referred to as `high $p_{{}_T}$ particle suppression', 
and is usually interpreted as the manifestation of the parton energy 
loss in the formed strongly interacting matter. Study on the energy 
dependence of the suppression phenomenon needed 
for its further understanding. The idea 
is that if the collision energy is low enough that this strongly 
interacting matter is not formed, the high transverse momentum 
particle suppression should disappear.

The particle suppression in a reaction A+B relative to that of p+p is 
measured by the nuclear modification factor 
$R^{BC}_{A+B}=\frac{1}{\left<N_{BC}\right>(A+B)}\cdot\frac{\mathrm{Yield}(A+B)}{\mathrm{Yield}(p+p)}$, 
where $\left<N_{BC}\right>(A+B)$ is the average number of binary collisions in the 
A+B reaction. (By the use of $\left<N_{BC}\right>$, it is implicitly assumed 
that the particle spectra, without any nuclear effects, should scale with the 
number of binary collisions.)

\begin{figure}[!ht]
\begin{center}
\includegraphics[width=14cm]{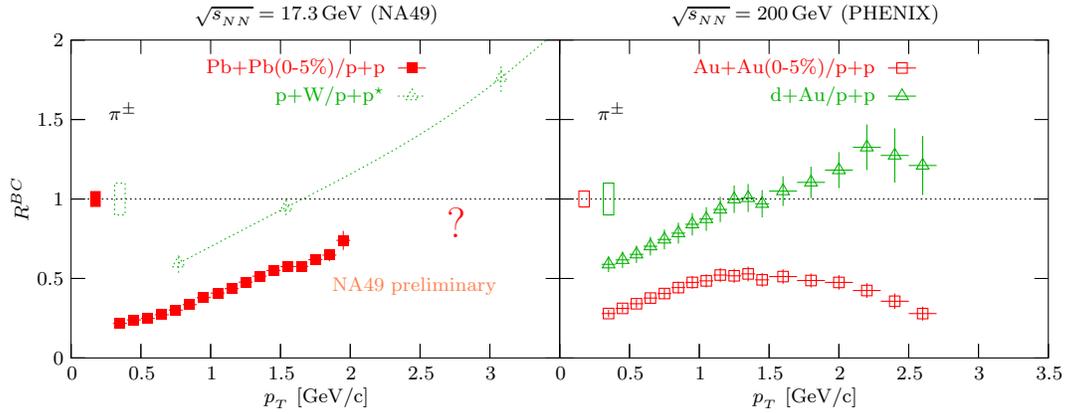}
\end{center}
\caption{Nuclear modification factor 
$R^{BC}$ for $\pi^{\pm}$ mesons in p+A and A+A interactions at the top 
SPS and RHIC energies. The rectangles 
show systematic uncertainty. The p+W/p+p curve was measured 
at $\sqrt{s_{{}_{NN}}}=19.4\,\mathrm{GeV}$ \cite{antreasyan}.}
\label{rbc}
\end{figure}

In Figure~\ref{rbc}, a preliminary $R_{AA}$ of $\pi^{\pm}$ production at 
$\sqrt{s_{{}_{NN}}}=17.3\,\mathrm{GeV}$ is shown, based on \cite{al} and \cite{pp} 
(left panel), and the published $R_{AA}$ result at $\sqrt{s_{{}_{NN}}}=200\,\mathrm{GeV}$ 
of \cite{phenix1,phenix2} is compared to it (right panel).
(The p+W/p+p curve at 
$\sqrt{s_{{}_{NN}}}=19.4\,\mathrm{GeV}$ was taken from \cite{antreasyan}.) 
It is seen that the NA49 curve stops at $p_{{}_T}\geq 2\,\mathrm{GeV/c}$, 
therefore the existence or non-existence of a suppression at higher $p_{{}_T}$ 
is not clear from the present data. The accessible $p_{{}_T}$ range 
is limited by the available p+p statistics.

\begin{figure}[!ht]
\begin{center}
\includegraphics[width=7.5cm]{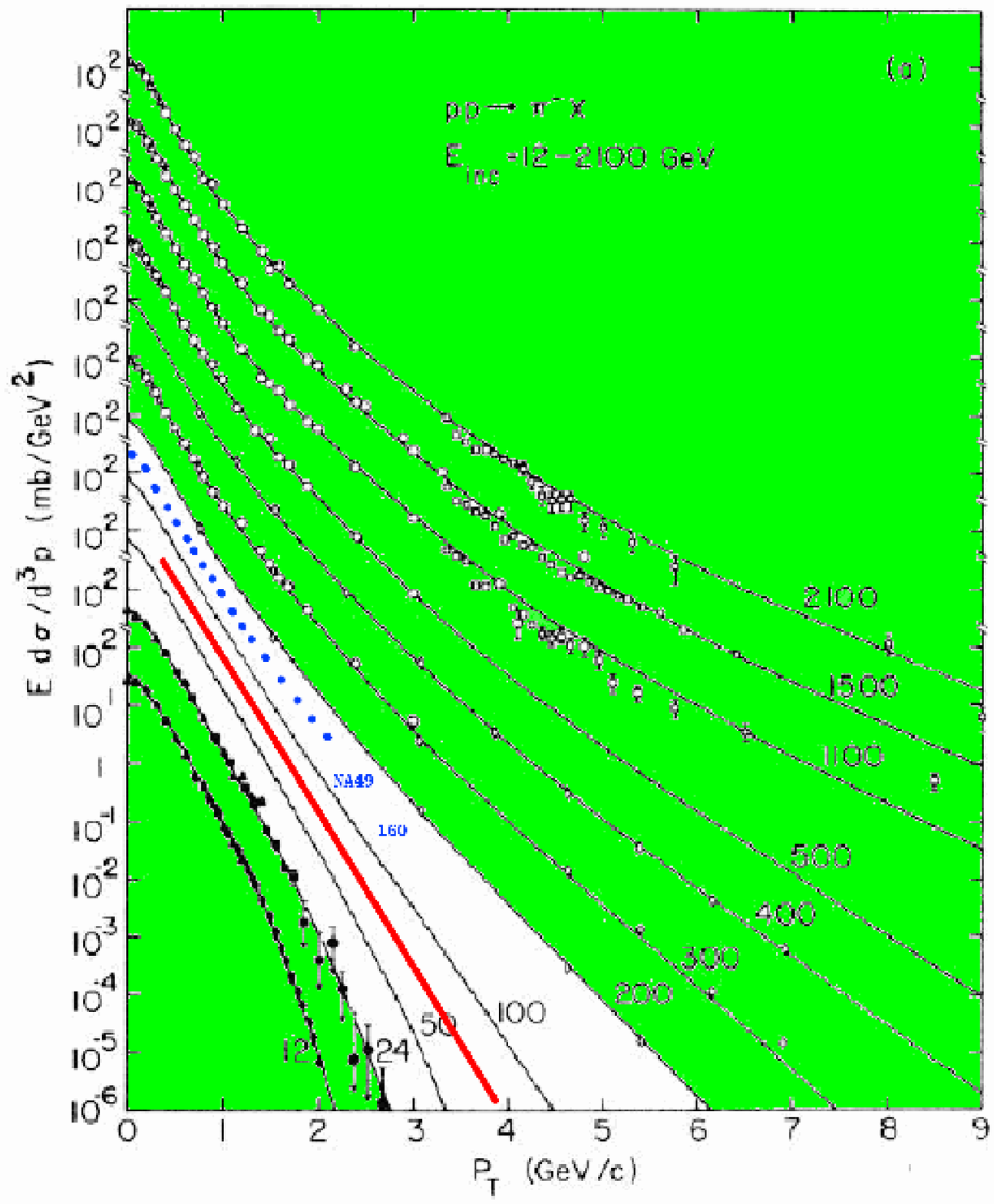}\includegraphics[width=7.5cm]{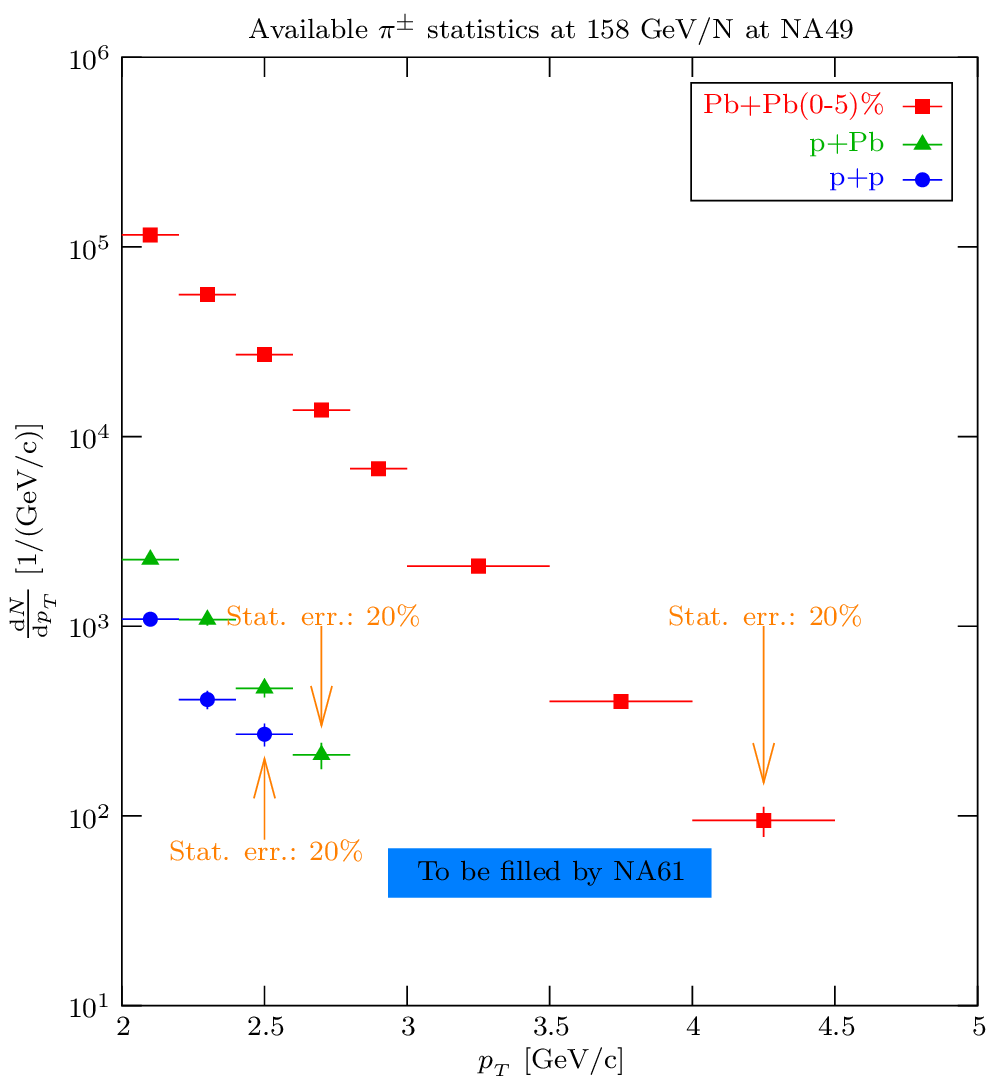}
\end{center}
\caption{Left: the available p+p data 
on $\pi^{\pm}$ production at our energy range, together with a 
parameterization (from \cite{beier}). The red line shows a change in spectrum shape. As can be seen: 
there are no data in the nearby energies, which makes the use of parameterization unsafe. 
Right: currently available high $p_{{}_T}$ $\pi^{\pm}$ statistics at midrapidity 
for Pb+Pb, p+Pb and p+p reactions at $\sqrt{s_{{}_{NN}}}=17.3\mathrm{GeV}$.}
\label{stat}
\end{figure}

Unfortunately, the available world-data on $\pi^{\pm}$ production in p+p 
does not cover our energy region. Furthermore, due to the closeness of 
the kinematic border of the available momentum space, the shape of the 
$p_{{}_T}$ spectra admits a rapid change from convex to concave 
at the SPS energy range, which makes an interpolation quite unsafe. 
This is shown in the left panel of Figure~\ref{stat}, 
together with a compilation of the world-data \cite{beier}. The 
red line marks the shape change from convex to concave with increasing 
beam energy. In the right panel of Figure~\ref{stat}, the currently available 
high $p_{{}_T}$ $\pi^{\pm}$ statistics of NA49 is shown 
for the 5\% most central Pb+Pb collisions as well as p+Pb and p+p interactions 
at $158A\,\mathrm{GeV}$ beam energy 
($\sqrt{s_{{}_{NN}}}=17.3\mathrm{GeV}$). It is seen, that the central Pb+Pb 
data extend up to $p_{{}_T}=4.5\,\mathrm{GeV}$, while the p+Pb and 
p+p spectra stop at about $2.5\,\mathrm{GeV/c}$. To estimate the necessary statistics, needed to be 
recorded by NA61, the Pb+Pb spectrum shape was used for extrapolation.

To accomplish the planned high statistics p+p and p+Pb measurements, the 
TPC readout system needs to be upgraded, to be able to record events at 
a higher rate.

\subsection{Reference spectra for T2K and cosmic-ray experiments}

\begin{figure}[!ht]
\begin{center}
\includegraphics[width=13cm]{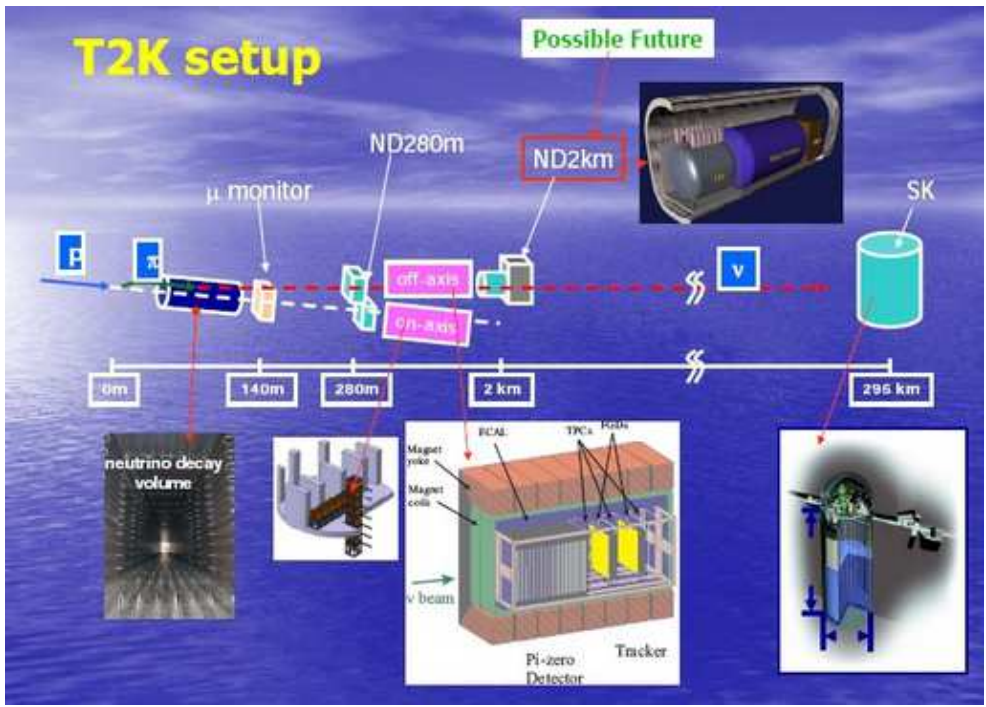}
\end{center}
\caption{The outline of the T2K experiment. The produced neutrino beam is 
detected at the Super Kamiokande, $295\,\mathrm{km}$ away from the neutrino 
beam production volume.}
\label{t2k}
\end{figure}

T2K \cite{t2k} is a neutrino oscillation experiment at JPARC, which produces 
neutrino beams indirectly by $p(30, 40, 50\,\mathrm{GeV})+C\rightarrow\pi,K+X$ reactions, 
and subsequent decay of pions and Kaons. The neutrinos 
are detected $295\,\mathrm{km}$ away, in the Super Kamiokande (SK) detector. 
The setup of the experiment is outlined in Figure~\ref{t2k}. 
The method of the $\nu_{\mu}\leftrightarrow\nu_{e}$ oscillation 
measurement is based on the so-called far-to-near ratio, $R$, of the neutrino 
fluxes in the far (SK) and the near (ND) detectors. To predict $R$ with a 
required precision, initial $\pi,K$ production cross-sections in p+C interactions 
at $30$, $40$ and $50\,\mathrm{GeV}$ should be measured. This will be done by 
NA61/SHINE experiment using a $1\,\mathrm{cm}$ ($2\%$ $\lambda_{\mathrm{int}}$) 
graphite target and a $90\,\mathrm{cm}$ long T2K replica target.

Cosmic-ray experiments are based on detection of extensive air showers 
by large lateral coverage ground detector arrays. The particle type and 
energy of the incident particle is then reconstructed by the simulation 
of the air shower. These simulations are most sensitive to $\mu^{\pm}$ 
production in cascades, which is related to the 
$p,\pi + C \rightarrow \pi,K + X$ production 
processes at SPS energies.\footnote{Such reactions occur when shower 
particles hit the nuclei of the air. 
A large fraction of muons detected by the ground detectors 
originates from decays of pions and Kaons produced in collisions 
at the CERN SPS energies.}
The $\mu^{\pm}$-s are produced by the weak decay of the $\pi,K$ particles. 
By the precise measurement of these production cross-sections, the model 
dependence of the extensive air shower simulations 
can be largely reduced. Some measurements of the NA61/SHINE are dedicated 
for this purpose.

For the T2K and cosmic-ray runs, the extension of the forward acceptance 
for particle identification is needed. Therefore, a new forward ToF wall for 
the NA61 setup is being built.

\section{NA61/SHINE, the upgraded NA49 detector}

\subsection{Detector setup}

\begin{figure}[!ht]
\begin{center}
\includegraphics[width=13cm]{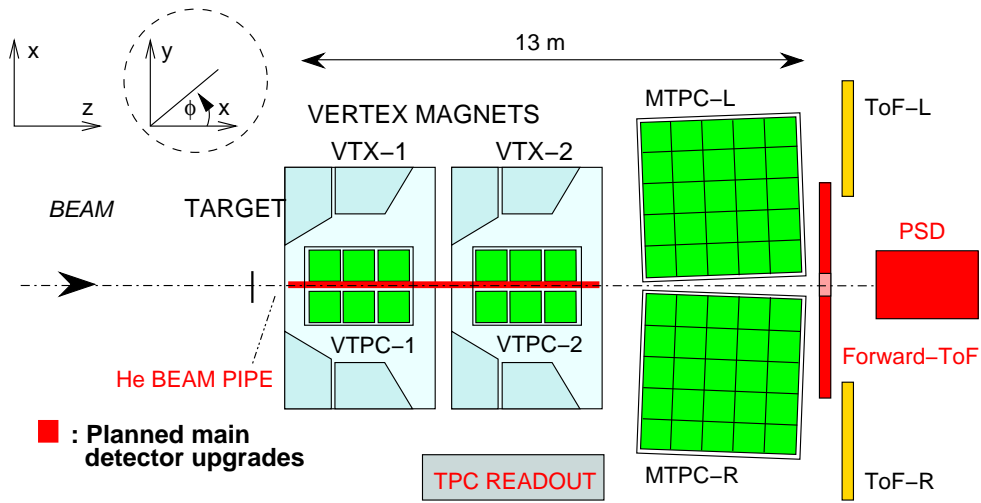}
\end{center}
\caption{The set-up of the NA61/SHINE experiment. The main detector 
detector upgrades are shown together with the original devices of NA49.}
\label{setup}
\end{figure}

The necessary detector upgrades, motivated by the physics goals, 
are shown in Figure~\ref{setup}. Due to its excellent performance, 
the NA49 TPC tracking system will be used in NA61. 
The most important performance parameters are listed below.\newline
\begin{tabular}{ll}
 $1.$ & Large acceptance: $\approx50\%$ at $p_{{}_T}\leq2.5\,\mathrm{GeV/c}$.\\
 $2.$ & Precise momentum measurement: $\Delta(p)/p^{2}\approx10^{-4}\mathrm(GeV/c)^{-1}$.\\
 $3.$ & High tracking efficiency: $\geq95\%$.\\
 $4.$ & Good particle identification: ToF resolution, $\sigma(t)\approx60\,\mathrm{ps}$, \\
     &  $\frac{\mathrm{d}E}{\mathrm{d}x}$ resolution, $\sigma(\frac{\mathrm{d}E}{\mathrm{d}x})\left/\frac{\mathrm{d}E}{\mathrm{d}x}\right.\approx5\%$, 
        invariant mass resolution, $\sigma(m)\approx5\,\mathrm{MeV}$.\\
\end{tabular}\\

The planned main upgrades are the followings.\newline
\begin{tabular}{ll}
 $1.$ & New spectator calorimeter, the PSD, instead of the VCAL. The 
       resolution to be achieved is \\
     & $\frac{\sigma(E)}{E}\approx\frac{0.5}{\sqrt{E/\mathrm{(1GeV)}}}$ ($\times5$ improvement), 
       and the uniformity to be achieved is \\
     & non-uniformity$<5\%$ ($\times20$ improvement). \\
     & (Motivated by the search for the critical point and onset of deconfinement.)\\
 $2.$ & Installation of a Helium beam-pipe to reduce $\delta$-electron contamination 
       in the sensitive \\
     & volumes ($\times10$ improvement). \\
     & (Motivated by the search for the critical point and onset of deconfinement.)\\
 $3.$ & Forward ToF wall (for particle identification in $p<3\,\mathrm{GeV/c}$ and $\theta<400\,\mathrm{mrad}$ region).\\
     & (Motivated by the T2K measurements.)\\
 $4.$ & New TPC readout with about $100\,\mathrm{Hz}$ readout frequency ($\times10$ improvement).\\
     & (Motivated by all physics goals.)\\
\end{tabular}

\subsection{Status and planning}

The beam request of NA61 is summarized in Table~\ref{beam}, together 
with the current recommendation and approval status assigned by the 
SPS Committee and CERN Research Board.

\begin{table}[!ht]
\begin{center}
\begin{tabular}{llllll}
 Beam & Energy [A GeV] & Year & Days & Physics & Status   \\
\hline
\hline
 $p$  & 30             & 2007 & 30   & T2K, CR & Approved \\
 $p$  & 30, 40, 50     & 2008 & 14   & T2K, CR & Recommended \\
 $\pi^{-}$  & 158, 350 & 2008 & 3    & CR      & Recommended \\
 $p$  & 158            & 2008 & 28   & high $p_{{}_T}$ & Recommended \\
 $S$  & 10, 20, 30, 40, 80, 158 & 2009 & 30    & CP\&OoD & Recommended \\
 $p$  & 10, 20, 30, 40, 80, 158 & 2009 & 30    & CP\&OoD & Recommended \\
 $In$ & 10, 20, 30, 40, 80, 158 & 2010 & 30    & CP\&OoD & To be discussed \\
 $p$  & 158            & 2010 & 30   & high $p_{{}_T}$ & To be discussed \\
 $C$  & 10, 20, 30, 40, 80, 158 & 2011 & 30    & CP\&OoD & To be discussed \\
 $p$  & 10, 20, 30, 40, 80, 158 & 2011 & 30    & CP\&OoD & To be discussed \\
\hline
\end{tabular}
\end{center}
\caption{The beam request of the NA61/SHINE experiment. Abbreviations: CP -- search for Critical Point; 
OoD -- study the Onset of Deconfinement; T2K -- supplementary spectra for the T2K experiment; 
CR -- measurements for cosmic-ray physics; high $p_{{}_T}$ -- p+p and p+Pb reference spectra 
for nuclear modification factors.}
\label{beam}
\end{table}

For the approval procedure of the NA61 experiment, the following documents are 
the most relevant: 
Expression of Interest \cite{eoi}, Letter of Intent \cite{loi}, Status Report 
\cite{sr}, Proposal \cite{prop}, Addendum-1 \cite{add1} and Addendeum-2 \cite{add2}.

\section{Summary}

The experiment NA61/SHINE has a great discovery potential to find 
the critical point of strongly interacting matter, if exists.

Important measurements of the nuclear modification factor 
at the top SPS energy can be performed and the system size dependence 
of the effects related to the onset of deconfinement can be studied.

The NA61 experiment at the CERN SPS is complementary to other 
projects on nucleus-nucleus collision  currently developed at 
FAIR, JINR, BNL and CERN LHC. It will provide the necessary 
input for the neutrino and cosmic-ray experiments at KEK, FZK.

\acknowledgments

This work was supported by the US Department of Energy, 
the Virtual Institute VI-146 of Helmholtz Gemeinschaft, Germany,
the Korea Science \& Engineering Foundation (R01-2005-000-10334-0),
the Hungarian Scientific Research Fund (OTKA 68506),
the Polish Ministry of Science and Higher Education (N N202 3956 33).

\end{document}